\documentclass[preprint,tightenlines,eqsecnum,floats,aps,amsmath,amssymb,nofootinbib,prd,showpacs]{revtex4}

\usepackage{amssymb}
\usepackage{stmaryrd}
\usepackage{amsmath}
\usepackage{amsfonts}
\usepackage{mathrsfs}
\usepackage{CJK}
\usepackage{amsmath,amssymb,amsfonts}
\usepackage{graphicx}
\def\be{\begin{equation}}
\def\ee{\end{equation}}
\def\ba{\begin{eqnarray}}
\def\ea{\end{eqnarray}}

\begin{document}
%\date\today
%\preprint{????}

\title{Theoretical constraints on properties of low-mass neutron stars from EOS of  inner crust matters}

\author{Yong-Mei Wen}
\affiliation{School of Physics, South China University of
Technology, Guangzhou 510641, P.R. China}

\author{ De-Hua Wen\footnote{Corresponding author. wendehua@scut.edu.cn}}
\affiliation{School of Physics, South China University of
Technology, Guangzhou 510641, P.R. China}

\date{\today}

%\date{\today}

\begin{abstract}

By employing four typical equation of states (EOSs) of the inner
crust matters, the    properties of   low-mass neutron stars  are
theoretically investigated. Based on the well-known fact that
there is a big gap between the neutron stars and white dwarfs in
the mass-radius sequence of compact stars, according to the
mass-radius relation of the four adopted EOSs,  we conclude a
rough forbidden-region for the central density and stellar radius
to form a compact stars, that is, there is no compact star in
nature having central density in the region from about
$10^{12}~\textrm{kg/m}^3$ to $10^{17}~\textrm{kg/m}^3$, and there
is also no compact star having a radius in the region from about
400 km to 2000 km. The  properties of the low-mass neutron stars
are also explored. It is shown  that for the stable neutron star
at minimum mass point, the stellar size (with radius $>$ 200 km)
is much larger than that of the normal neutron stars (with radius
about 10 km), and there is a compact $'$core$'$ concentrated about
95\% stellar mass in the inner core with radius within 13 km and
density higher than the neutron-drip point ($4.3\times 10^{14}
~\textrm{kg/m}^{3}$). This property is totally different from that
of the normal neutron stars and white dwarfs. For a stable neutron
star with stellar mass near the minimum mass, the Keplerian period
is several hundred millisecond, the moment of inertia is in an
order of $10^{37} kg\cdot m^{2}$, and the surface gravitational
redshift is in an order of $10^{-4}$.
\end{abstract}

\pacs{97.20.RP;71.10.-w;04.40.Dg}

\maketitle

\section{Introduction}

 Neutron star
 is one of  the hottest  objects of study during the past decades,
 as neutron star provides a gold mine for the  fundamental physics, including nuclear physics,
astrophysics,  particle physics and general relativity. Meanwhile,
the nature of matter at super high densities is one of the great
unsolved problems in modern science, and  it is believed that
    neutron star is the natural and
irreplaceable laboratory for investigating the super dense matters
\cite{Lattimer2004,Piekarewicz2014,Lattimer2016,Watts2016}.
Therefore, among the multitudinous studies on the neutron stars,
one important direction is to survey the constraints on the
equation of state (EOS) of the dense matters based on the accurate
and reliable observations on neutron stars, such as the fastest
observed spin frequency (716Hz)\cite{Hessels2006} and the observed
massive masses ($\sim 2.0 M_{\odot}$)
\cite{Demorest2010,Antoniadis2013}, where  both of them   come
from radio pulsars. In this direction,  it is worth looking
forward to obtaining more strict constraint on the equation of
state of dense matter by measuring the radius of neutron stars  to
accuracies of a few percent by the next generation of hard x-ray
timing instruments \cite{Watts2016}. Another interesting direction
is the theoretical researches on the EOS of the dense matter based
on the experiments in terrestrial laboratory and further
 investigations on the structure of neutron star by using the EOS,
which can provide a useful guide in primary understanding the
properties of pulsars \cite{Lattimer2004,Li2008,Lattimer2016}.

Presently,   two characteristic parameters of the pulsars (neutron
stars) have been observed very accurately, that is, the spin
frequencies (or periods) and the stellar masses. More than 2400
pulsars are now observed \cite{Manchester2015}.  As most of the
pulsars are identified  by their pulses (it is believed that the
beams of radiation waves come from the magnetic pole region with
one pulse per rotation. it is worth noting that in some cases, two
pulses per rotation may be detected if the star¡¯s magnetic and
rotation axes are nearly orthogonal \cite{Kaspi2016}), this means
that the spin frequencies for most of the known neutron stars have
been accurately observed. Typical observed spin period is a few
hundred millisecond (ms). Presently, the known spin periods of
neutron stars are in a region from 1.4 ms to 12s
\cite{Hessels2006,Manchester2015,Kaspi2016}, and based on which
the neutron stars are classified as two groups: normal pulsars
with spin periods P$>$30 ms and millisecond pulsars (MSPs) with
P$<$30 ms, where MSPs comprise about 15\%  of the known pulsar
population \cite{Manchester2015}.  It is easy to understand that
the massive-mass and small-size neutron star can support a very
rapid spin frequency as  the centrifugal force can be balanced by
the super strong gravity. Theoretical investigations show that a
high-mass neutron star (with mass $\sim 2 M_{\odot}$) can support
a sub-millisecond spin period \cite{Krastev2008,Wen2011}; and even
a low-mass neutron star still can support a relative rapid spin
period (for example, a neutron star with $M \sim 0.5 M_{\odot}$
can support a period $\sim$ 2ms )
\cite{Cook1994,Krastev2008,Wen2011}. On the other hand, being
different from the spin period observation, there are only few
neutron stars having accurate stellar-mass measurements. Up to
date, the number of  reliable and precise mass-measured neutron
stars is only 32 \cite{Antoniadis2016}. These samples  still can
not be used as an observational evidence to constrain the upper or
lower mass-limit for neutron stars. Among these accurate
mass-measured neutron stars, J0348+0432 has the highest mass ($M=
2.01\pm 0.04 M_{\odot}$), and J0453+1559(c) has the lowest mass
($M= 1.174\pm 0.004 M_{\odot}$)\cite{Antoniadis2016,Martinez2015}.
In addition, the fraction of the massive NSs (with $ M >
1.8M_{\odot}$) is about 20$\%$ of the observed population
\cite{Antoniadis2016}.

 There is an
untended inconsistency between the  theoretical results and
observations: the lower mass limit of the cold non-rotating
neutron star is about $ 0.1 M_{\odot}$ in theory while the
observed lowest mass of neutron star is $ 1.174\pm 0.004
M_{\odot}$ \cite{Haensel2002,Martinez2015}. And up to date all of
the observations, including the spin frequencies
 and the stellar masses, still can not confirm or rule out the existence of the  low-mass neutron
 stars in the universe. Theoretically, it is found that for the newly born proton-neutron stars, because
of the larger  thermal and neutrino-trapping effects, the minimum
mass limit of neutron star is expected to increase  to
0.9$\sim$1.1 $M_{\odot}$ \cite{Goussard1998,Strobel1999}. This
means  that a low-mass neutron star can not be produced directly
by the supernova explosion. As we know, among the 2500 observed
neutron stars there are only about 30 neutron stars having
accurate measured-masses, and the existing knowledge still cannot
rule out the possibility that there may be some low-mass neutron
star in the nature. But how to form such a low-mass neutron star
is still an unsolved mystery. We live in  hope, there may be some
unknown mechanism to form a low-mass neutron star. In fact, as an
attempt, one possible mechanism has been proposed about twenty
years ago \cite{Sumiyoshi1998,Haensel2002}. That is, in a binary
pulsar system, if the distance between two stars is sufficiently
small, the star with less mass and larger size will lose mass due
to the gravitational pull by the companion star with high mass and
small size. This is a
 self-accelerating process, because decrease of mass leads to
increase of neutron star size, and vice versa. And this process
makes it even more accessible for the mass loss until the less
mass star reaches its minimum mass limit ($M_{min}$).  Numerical
simulations suggest that when the less mass neutron star crosses
the $M_{min}$, it will undergo an explosion
\cite{Sumiyoshi1998,Haensel2002}.

 One inspiring news for the neutron star observation is that the
Five-hundred-meter Aperture Spherical radio Telescope (FAST), the
largest single dish radio telescope in the world, has been
completed by Sept. 2, 2016 in China. FAST's high instantaneous
sensitivity provides the potential capacity to discover thousands
of pulsars in the near future, which is far more than the number
of the observed pulsars \cite{fast}. Thus, it is hopeful that FAST
may increase substantially the number of the
accurate-mass-measured neutron stars and even find the low-mass
neutron star in the universe. On the other hand,
 Most of the theoretical
researches focused on the normal neutron stars (with mass $ \sim
1.4 M_{\odot}$) or the massive neutron stars (with mass $ \sim 2
M_{\odot}$), but few of them concentrated on the low-mass neutron
stars (with mass $ < 0.5 M_{\odot}$)
\cite{Lattimer2004,Li2008,Antoniadis2013,Lattimer2016,Antoniadis2016}.
Since there are possibilities in theory to exist the low-mass
neutron stars, it is interesting to investigate them further,
especially the
 observable quantities, which may provide useful guide for the future observation on the low-mass
neutron stars.
 Therefore, in this work we will focus on the
low-mass neutron stars, especially for those above the minimum
mass limit. Moreover, if the low-mass neutron star is observed in
the near future, it is also interesting to study the relations
between the EOS of the inner crust and the global properties of
the low-mass neutron star, as the low-mass neutron stars comprise
the information of inner crust matter, in which the compositions
and equation of states are  not well assured yet. As we know, all
the observed pulsars have fast spin frequencies. When we
theoretically investigate the property of neutron star, the
rotation is an important parameter. But  to the cases with
rotation frequency far below the Kepler frequency, where the
rotation effect becomes insignificant, people often deal with the
neutron stars as non-rotating stars. As a first step, here we also
suppose that the low-mass neutron star considered here rotates far
below its Kepler frequency, and thus can be treated as
non-rotating star.

This paper is organized as follows: after a short introduction, we
present the EOS of the low-mass neutron star matter in Sec. II.
The structure equations and quantities of the non-rotating neutron
stars are introduced in Sec. III. The detailed properties of the
low-mass neutron stars and discussions are shown in Sec. IV,
Conclusions and outlooks are given in the last section.

\section{The EOS of the low-mass neutron star matter}

Normally, a neutron star is believed to be composed of three main
parts: (1)  the outer crust, with density ranging from about $10
~\textrm{kg/m}^{3}$ to the density of neutron-drip point
$4.3\times 10^{14} ~\textrm{kg/m}^{3}$ \cite{Baym1971a}; (2) the
inner crust, with density ranging from  $4.3\times 10^{14}
~\textrm{kg/m}^{3}$ to the density of crust-core transition, which
is about half of saturation nuclear density $\rho_{s}$, but not
assured yet; (3) the uniform inner core, with density up to
several times of $\rho_{s}$ at the stellar center. The equation of
state (EOS) of neutron star matter is a basic input for
theoretical investigation on the construction of neutron star.

 In the outer crust, at densities below about $10^{7} ~\textrm{kg/m}^{3}$ a fraction
 of electrons are bound to the nuclei; while
 at densities between
  $10^{7} ~\textrm{kg/m}^{3}$ $\sim$ $4.3\times 10^{14} ~\textrm{kg/m}^{3}$
  the electrons are  free and
soon become relativistic as the density increases
\cite{Baym1971a}. The EOS of the ground state of the outer crust
can be well determined by using the experimental masses of
neutron-rich nuclei \cite{Haensel1994}. Thus there is no distinct
difference among the different models for the outer crust. In
theoretical investigation on the neutron star structure, many of
the researchers adopted the BPS models \cite{Baym1971a,
Lattimer2016} as the EOS of outer crust matter. Another improved
EOS for the outer crust is established by Haensel and Pichon based
on the  progress in the experimental determination of masses of
neutron rich nuclei \cite{Haensel1994}.

\begin{figure}
\centering
\includegraphics [width=0.6\textwidth]{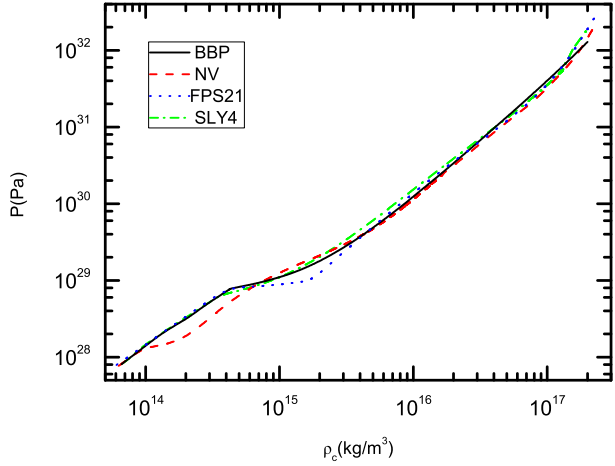}
\caption{ \label{1} The equation of states (EOSs) for the inner
crust. }
\end{figure}

The inner  crust of neutron stars comprises the region from
neutron-drip surface  up to the crust-core transition edge, inside
which the dense matter are believed to melt into the uniform
liquid core.  Above the neutron drip point, the energetic neutrons
are no longer bound by the nuclei and the dripped neutrons form a
free neutron gas. Thus, the matter in the inner crust is believed
to be composed of nuclei, neutrons and electrons under the
conditions of $\beta$ stability and charge neutrality. The
properties of nuclei in the inner crust matter are   expected
being very different from those of terrestrial nuclei because
their properties are influenced by the gas of dripped neutrons
\cite{Negele1973}. Therefore, the EOS of the inner crust matter is
still theoretical model dependent. In this work, four
representative EOSs for the  inner crust matter will be employed
to investigate the low-mass neutron stars. They are the BBP EOS
\cite{Baym1971a, Baym1971}, the NV EOS \cite{Negele1973}, the
FPS21 EOS \cite{Pethick1995}, and the SLY4 EOS \cite{Douchin2001}.
The BBP EOS describes the nuclei by a compressible liquid-drop
model to take  into account the effects of the free neutrons which
exert a pressure on the surface of the nuclei and thus lower the
nuclear surface energy. Another important physical factor first
taking into account in this model is the attractive Coulomb
interaction between nuclei (that is, the lattice binding energy),
which is very important in determining the size of nuclei. The NV
EOS is constructed  by Negele and Vautherin in 1973, which  is the
first group to use the Quantum calculations investigating the
matter of the inner crust \cite{Negele1973}. They determined the
structure of the inner crust by minimizing the total energy per
nucleon in a Wigner-Seitz sphere, and   treating the electrons as
a relativistic Fermi gas. The FPS21 EOS employs a generalized type
of Skyrme interaction, which fits well   for both the nuclear and
neutron-matter calculations of Friedman and Pandharipande
\cite{Pethick1995,Friedman1981}.  The SLY4 EOS is based on the SLy
(Skyrme Lyon) effective nucleon-nucleon interactions with a new
set of improved parameters which can reproduce the new
experimental data very well \cite{Douchin2001,Chabanat1998}. For a
detailed review for the physics of neutron star crusts, please
refer to the literature \cite{Chamel2008}. The EOSs of  inner
crust adopted in this work are presented in Fig. \ref{1}. It is
shown that the main disparity among these EOSs is in the density
region from about $1.0\times 10^{14}~\textrm{kg.m}^{-3}$ to
$5\times 10^{15}~\textrm{kg.m}^{-3}$, where   FPS21 EOS, SLY4 EOS
and BBP EOS are stiffer than NV EOS in the density region from
about $1.0\times 10^{14}~\textrm{kg.m}^{-3}$ to $8\times
10^{14}~\textrm{kg.m}^{-3}$, and SLY4 EOS, BBP EOS and NV EOS are
stiffer than FPS21 EOS in the density region from about $8\times
10^{14}~\textrm{kg.m}^{-3}$  to $5\times
10^{15}~\textrm{kg.m}^{-3}$.

\begin{figure}
\centering
\includegraphics [width=0.6\textwidth]{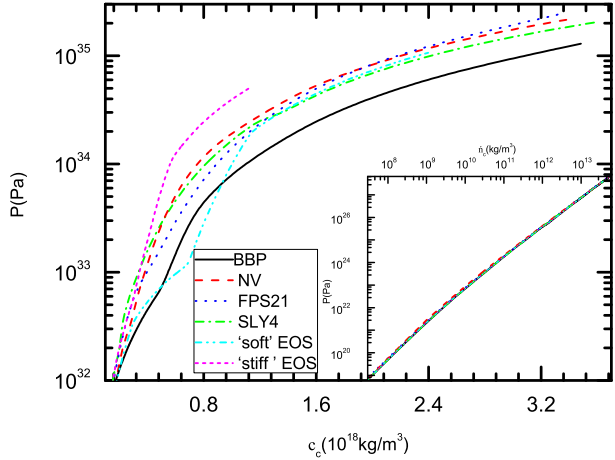}
\caption{\label{2} The EOSs for the inner core, where the inset
describes the EOSs for the outer crust. }
\end{figure}

In order to investigate the whole sequences of the neutron stars
described by different EOSs models, we also calculate the whole
mass-radius relations of the compact stars from the white dwarf
models up to the upper mass limit of neutron stars, where the EOSs
of the inner crust are smoothly joined to the EOSs of the uniform
liquid inner core and the EOSs of the outer crust  based on the
following literatures (For
  simplifying  the expression of EOSs, in the following
part of this paper,  the EOSs combining the outer crust and the
inner core are also denoted by above abbreviation of EOSs for the
inner crust): (1) For the BBP EOS, the outer crust is connected by
the BPS EOS,  the inner core is combined by the EOS constructed by
Pandharipande \cite{Pandharipande1971}, where the parameters of
the  crust-core transition are $\rho_{t}=2.0\times
10^{17}\textrm{kg}\cdot \textrm{m}^{-3}$, $n_{t}=1.18\times
10^{44}\textrm{m}^{-3}$, $p_{t}=1.29\times 10^{32}\textrm{pa}$
\cite{Baym1971a}, and the whole data of the joined EOS are taken
from Ref. \cite{Baym1971a}; (2) For the NV EOS, the EOS of the
outer crust is taken from the work of Harrison and Wheeler
\cite{Harrison1965}, the DBHF EOS is employed to describe the
inner core \cite{Krastev2006},  where the parameters of the
crust-core transition are   $\rho_{t}=2.0\times
10^{17}\textrm{kg}\cdot \textrm{m}^{-3}$, $n_{t}=1.17\times
10^{44}\textrm{m}^{-3}$, $p_{t}=1.37\times 10^{32}\textrm{pa}$
\cite{Negele1973,Sammarruca2009}, and the whole data of the joined
EOS are taken from Ref. \cite{Sammarruca2009}; (3) For the FPS21
EOS, the EOS of the outer crust is constructed by Haensel and
Pichon \cite{Haensel1994}, the inner core is joined by the famous
APR EOS \cite{Akmal1998},  where the parameters of the  crust-core
transition are $\rho_{t}= 1.8\times 10^{17}\textrm{kg}\cdot
\textrm{m}^{-3}$, $n_{t}=0.98\times 10^{44}\textrm{m}^{-3}$,
$p_{t}=0.8\times 10^{32}\textrm{pa}$ \cite{Pethick1995}, and the
whole data of the joined EOS are kindly provided by Mr. Krastev
\cite{Krastev2008}. (4) For the SLY4 EOS, the outer crust is also
connected by the model of Haensel and Pichon \cite{Haensel1994},
and the inner core is described by the SLy4 model
\cite{Chabanat1998}, where the parameters of the  crust-core
transition are $\rho_{t}=1.28\times 10^{17}\textrm{kg}\cdot
\textrm{m}^{-3}$, $n_{t}=0.76\times 10^{44}\textrm{m}^{-3}$,
$p_{t}=0.54\times 10^{32}\textrm{pa}$ \cite{Douchin2001}, and the
whole data of the joined EOS are taken from Ref.
\cite{Douchin2001}.  In fact, the transition region from the
highly ordered crystal to the uniform liquid core is complicated
and is poorly understood, and up to date, the transition density
is still very divergent
\cite{Baym1971,Negele1973,Douchin2001,Xu2009,Ducoin2011,Piekarewicz2014}.
Fortunately, the position of the crust-core transition edge is
mainly responsible for such as the glitches, and does not directly
effect the mass-radius relation and the lower mass limit for a
certain EOS. By the way, people often use $'$softness$'$ or
$'$stiffness$'$ to describe an EOS, where a softer EOS means that
at a fixed density  the pressure is relative lower, while a
stiffer EOS corresponds to a higher pressure
\cite{Ozel2006,Demorest2010,Hebeler2013,Lattimer2016}. In fact,
the softness and stiffness for an EOS is relative. A clear express
should include the density domain and the comparison object. For
example, the NV EOS is stiffer than most of the other considered
EOSs at higher density region (as shown in Fig. \ref{2}), while it
is softer than other EOSs in density region $1.0\times
10^{14}~\textrm{kg.m}^{-3}\sim 8\times 10^{14}~\textrm{kg.m}^{-3}$
and $10^{16}~\textrm{kg.m}^{-3}\sim 10^{17}~\textrm{kg.m}^{-3}$
(as shown in Fig.  \ref{1}). In this work, unless otherwise
specified,  when we talk about the softness (or stiffness) of an
EOS, the density domain is in $10^{14}~\textrm{kg.m}^{-3}\sim
10^{17}~\textrm{kg.m}^{-3}$, that is, for the matter in inner
crust; and the comparison objects are the four considered EOSs.

The EOSs for the inner core are presented in Fig. \ref{2}, where
the EOSs of the outer crust is shown in the inset. It is shown
that for all of the adopted EOSs, the EOSs of the outer crust
matter are similar. And for the inner core, BBP EOS is relatively
softer. For comparing with the  new progress of EOSs, we also plot
the high density part of $'$soft$'$ and $'$stiff$'$ EOSs of Ref.
\cite{Hebeler2013} in Fig. \ref{2}, where the  crust are connected
by the BBP EOS ( denote as BPS in the Ref. \cite{Hebeler2013}).
 It is shown that the $'$soft$'$ EOS is still
stiffer than BBP EOS at higher density ( $> 3\rho_{0}$). In order
to express conformable with the reference and also do not make
confusion in this work, here we add an apostrophe to denote these
two EOSs.

\section{Structure equations of non-rotating neutron stars}

Supposing that the matter of a neutron star can be treated as
perfect fluid, then its energy-momentum tensor can be described as
  \begin{equation}
T^{\alpha\beta}=pg^{\alpha\beta}+(p+\rho)u^{\alpha}u^{\beta},
\end{equation}
where $\alpha,\beta=0,1,2,3$,  $u^{\alpha}$ is the four-velocity
satisfying $u^{\alpha}u_{\alpha}=-1$, $\rho$ is the density and
$p$ is the pressure. Unless otherwise noted, we use geometrical
unit ($G$ = $c$ = 1). A non-rotating spherically symmetric neutron
star is described by the following metric
\begin{equation}
-ds^{2}=-e^{2\Phi}dt^{2}+e^{2\Lambda}dr^{2}+r^{2}(d\theta^{2}+\textrm{sin}^{2}\theta
d\phi^{2}),
\end{equation}
where $\Phi$ and $\Lambda$  are the functions of radius $r$ only.
According to the Einstein field equation
 \begin{equation}
R^{\alpha\beta}-\frac{1}{2}g^{\alpha\beta}R=8\pi T^{\alpha\beta}
\end{equation}
and combing it with  Eqs.(2) and (3),  one can obtain the
non-rotating neutron star  structure equations, namely, the
familiar Tolman-Oppenheimer-Volkoff (TOV) equations\cite{s23,s24}
  \begin{equation}
\frac{dp}{dr}=-\frac{(p+\rho)[m(r)+4\pi r^{3}p]}{r[r-2m(r)]},
\end{equation}
  where
\begin{equation}
m(r)=\int_{0}^{r}4\pi r^{2}\rho dr.
\end{equation}
 The EOS of the neutron star
matter is a necessary input to solve the TOV equations. We can
integrate outwards from the origin ($r = 0, \rho=\rho_{c}$) to the
point $r = R$, where the pressure becomes zero. This point defines
$R$ as the radius of the star and $M = M(R)$   as the total mass
of the star.

Except for the mass and the radius, there are several other
interesting global
      quantities characterizing the neutron stars, such as the moment of inertia,
      the redshift, etc. Because of the strong gravity in neutron
      star, these quantities are normally calculated in general
      relativistic theory. For the non-rotating neutron star, its moment of inertia can be calculated by \cite{Wen2011,Morrison2004}
 \begin{equation}
I=\frac{8\pi}{3}\int_{0}^{R}e^{-\Phi}\frac{p+\rho}{\sqrt{1-\frac{2m(r)}{r}}}r^{4}dr,
\end{equation}
 where the rotational effect in the original equation is
 neglected.  As we only pay attention to the low-mass
 neutron star in this work, where the compactness (${2M}/{R}$) is very small (in an order of $10^{-2}$ to $10^{-3}$, even
 reduced
to $10^{-4}$ at the minimum mass point), far smaller
 than that of a normal neutron star,  so it is enough to calculate the moment of inertia in Newtonian gravity, that is
\begin{equation}\label{I}
I=\frac{8\pi}{3}\int_0^R\rho {r^4dr}.
\end{equation}

The surface gravitational redshift can be simply calculated by
 \ba Z=(1-\frac{2M}{R})^{-1/2}-1\label{Z} \ea.

For a star formed by perfect fluid, there exists a maximal
rotating frequency at which there   comes into being a balance of
gravitational and centrifugal forces at the star's equator.
Exceeding the maximal frequency, the star will engender mass
shedding. This  maximal frequency is named as Kepler frequency.
Similarly, because of the relative weaker gravity of the low-mass
neutron star, we also calculate the Kepler frequency here in the
Newtonian gravity frame, that is
 \ba
P_{K}=2\pi\sqrt\frac{R^3}{M}\label{p}.
\ea

\section{The properties of the low-mass neutron stars}

\begin{figure}
\centering
\includegraphics [width=0.6\textwidth]{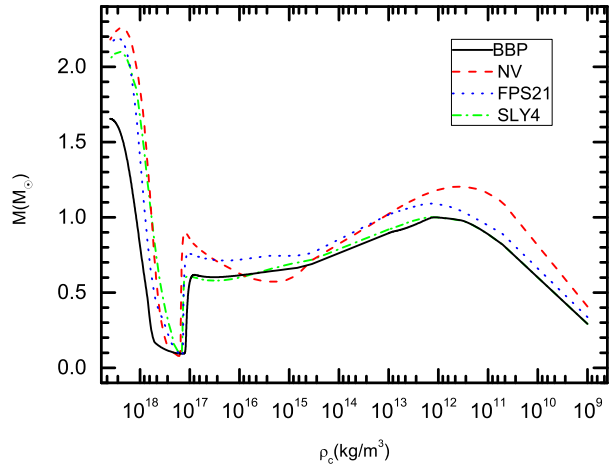}
\caption{ \label{3} The stellar masses as a function of the
central densities for the four adopted EOSs, where the central
density ranges from $ 10^9~kg/m^3$ to $4\times10^{18}~kg/m^3$ and
the star sequence covers from  white dwarfs to neutron stars. }
\end{figure}

Theoretical investigation on the minimum mass for neutron stars
has a long history. The first estimate of minimum mass $M_{min}$,
which is about 0.17 $M_{\odot}$ without taking into account the
nuclear interactions, was obtained by Oppenheimer \& Serber
\cite{Oppenheimer1938}. Updated research shows that the minimum
mass is  around  0.09 $M_{\odot}$ and has a weakly EOS dependence
\cite{Douchin2001}. For a low-mass neutron star in binary system,
if it loses its mass below the minimum mass $M_{min}$   through
gravitational pull by the companion star, an explosion will be
undergone \cite{Sumiyoshi1998,Haensel2002}. On the other hand, for
the
  stars sequence in the $M-\rho_{c}$ curve with central densities $\rho_{c}$ smaller than that of  $M_{min}$, where $ dM/d\rho_{c}<0 $ , as shown in Fig. \ref{3},
equilibrium configuration will also do not exist, as it will
become unstable with respect to small radial perturbations
\cite{Shapiro1983,Glendenning2000}. Thus, we only need to care for
the stable low-mass neutron stars with mass above the minimum mass
$M_{min}$.

\begin{figure}
\centering
\includegraphics [width=0.6\textwidth]{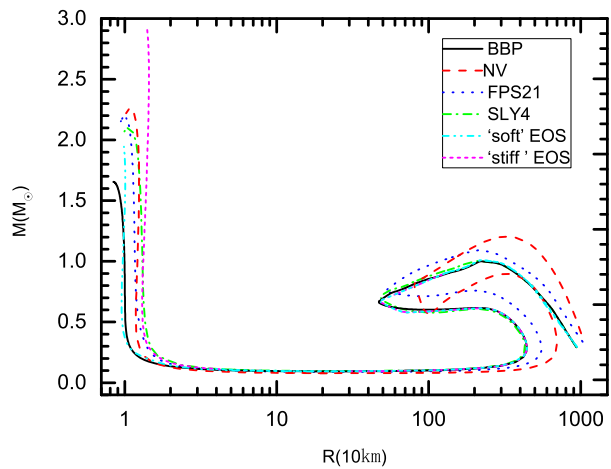}
\caption{ \label{4} The mass-radius relations for the four adopted
EOSs. For comparison, the mass-radius relations for the $'$soft$'$
and $'$stiff$'$ EOSs of Ref. \cite{Hebeler2013} are also plotted.
}
\end{figure}

\begin{figure}
\centering
\includegraphics [width=0.6\textwidth]{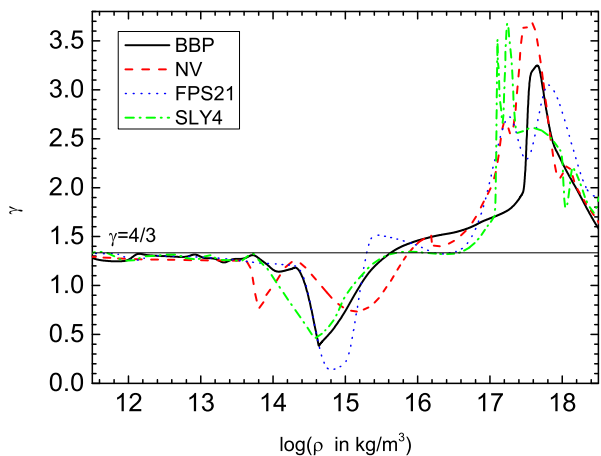}
\caption{ \label{index} The  adiabatic indices in  the density
ranging from   $3\times 10^{11} ~\textrm{kg/m}^{3}$ to $ 3\times
10^{18} ~\textrm{kg/m}^{3}$ as a function of the density for the
four adopted EOSs. }
\end{figure}

In order to find the common characters and also show the effect of
the EOSs on the low-mass neutron stars, we choose four typical
EOSs of the inner crust, as have been described in Sec. II, to
investigate the  properties and structures of the low-mass neutron
stars. For convenient to understand the whole star sequence and
where  the low-mass neutron stars located, we plot the
$M-\rho_{c}$ relations in Fig. \ref{3} and $M-R$ relations in Fig.
\ref{4} in a   wide central density   region (from $ 10^9~kg/m^3$
to $4\times10^{18}~kg/m^3$ ). And in order to compare our results
with that of the new progress of  EOS, the $M-R$ relations of the
$'$soft$'$ and $'$stiff$'$ EOSs of Ref. \cite{Hebeler2013} are
also plotted in Fig. \ref{4}. It is shown that there is no
distinct difference between the $'$new$'$ EOSs and the $'$old$'$
EOS at the low-mass region for stable neutron stars.

\begin{figure}
\centering
\includegraphics [width=0.6\textwidth]{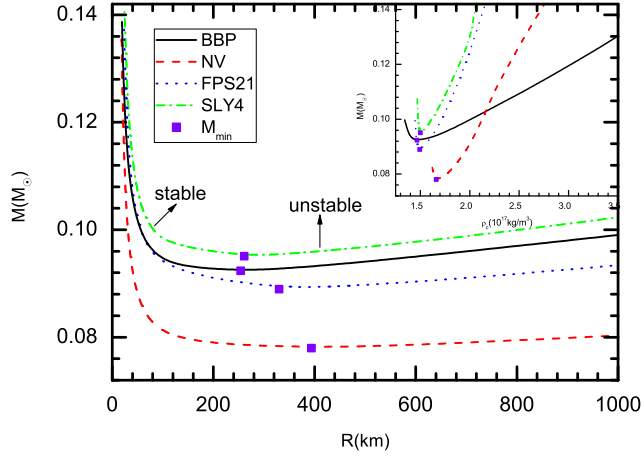}
\caption{\label{5} The mass-radius relations for the low-mass
neutron stars, where the inset is the masses as a function of the
central densities, and the small square on the line denotes the
$M_{min}$ point. }
\end{figure}

As we know,  white dwarfs are the lowest density sequence of
stable compact stars, where the central densities take a range
from about $ 10^{9}~kg/m^3$ to $10^{12}~kg/m^3$, as shown in Fig.
\ref{3}. From a microscopic point of view, in a stable white
dwarfs the inward gravity is balanced by the pressure of
degenerate electrons.  As the mass of white dwarf increases, the
central density and thus the Fermi energy of electrons also rise.
When the Fermi energy of electron reaches a point at which the
energetical electrons can be captured by protons through inverse
beta decay, the pressure of the relativistic electrons stops
increasing along with the density increasing, while the neutrons
produced by the inverse beta decay are still bound in the
neutron-rich nuclei until the neutron-drip point ( $4.3\times
10^{14} ~\textrm{kg/m}^{3}$) reaches. Thus there is not enough
pressure to support the strong gravity when the central density of
a star becomes higher than that of the white dwarf with
Chandrasekhar mass limit ($1.44 M_{\odot}$) until the pressure of
degenerate neutrons becomes stronger enough to balance the gravity
\cite{Glendenning2000}. From a macroscopic point of view,  if the
EOS of a star can be approximated by
 \ba p=K\rho^{\gamma} \ea
with $K$   a constant coefficient and $\gamma$   the adiabatic
index,
 then the mass and radius have the relationship
\cite{Glendenning2000}
 \ba R \sim M^{(\gamma-2)/(3\gamma-4)}. \ea
In order to satisfy the necessary stable condition
\cite{Glendenning2000} \ba \frac{\partial M(\rho_{c})}{\partial
\rho_{c}}>0,  \ea one must have
 \ba\gamma>4/3. \ea
While for the ultra-relativistic electron (with density
$\rho>>\rho_{c}\sim 10^{9}\mu $ kg$\cdot$ m$^{-3}$, where $\mu$ is
the number of nucleons per electron \cite{Weinberg1972}), its
polytropic index $\gamma=4/3$. This polytropic index sets the
famous Chandrasekhar mass limit for white dwarfs. Beyond the
density about $10^{12}~$kg$\cdot$m$^{-3}$  neutronization will
destabilize the white dwarfs, until the   pressure of degenerate
neutrons dominates the balance against inward gravity. As an
approximative and qualitative discussion for the star's stability,
we suppose that the tabulated EOSs  also can be approximatively
expressed by  $ p=K\rho^{\gamma} $ and the adiabatic index can be
calculated through $\gamma=\frac{d \ln(p)}{d \ln(\rho)}$. In Fig.
\ref{index}, we plot the adiabatic indices of the four adopted
EOSs in  the density ranging from $5\times 10^{11}
~$kg$\cdot$m$^{-3}$ to $ 10^{17}$ kg$\cdot$m$^{-3}$. It is shown
that in the density ranging from $\sim 10^{12} $ kg$\cdot$m$^{-3}$
to $ \sim 10^{15} $kg$\cdot$m$^{-3}$, the adiabatic indices of the
four  EOSs are less than 4/3. Obviously, a central density in this
density region cannot support a stable star. As to the stars with
central densities ranging from $\sim 10^{15} ~$kg$\cdot$m$^{-3}$
to $ \sim 10^{17} ~$kg$\cdot$m$^{-3}$, though the adiabatic
indices in this density region start to become larger than 4/3,
the star sequences of the EOSs  still cannot obey the
  necessary stable condition
$ \frac{\partial M(\rho_{c})}{\partial \rho_{c}}>0 $, as shown in
Fig. \ref{3}. That is, in this density region the stars are still
unstable (We guess that one possible reason for this case may be
that we cannot approximate the tabulated EOSs by a single
polytropic form in such a wide density region).
 Because of the reason discussed above, we conclude that between the white dwarfs and neutron stars there
 exists
a wide forbidden region for the central densities (from about
$10^{12}~$kg$\cdot$m$^{-3}$ to $10^{17}~$kg$\cdot$m$^{-3}$) to
form a compact star, as shown in Fig. \ref{3}. Correspondingly,
for the same reason, we can conclude a rough size forbidden-region
for the compact stars according to the four adopted EOSs in this
work, that is, there is no compact star having a radius in the
region from about 400 km to 2000 km,  as shown in Fig. \ref{4}.
 %One possible reason for $\sim 10^{15} ~\textrm{kg/m}^{3}$ to $ \sim 10^{17} ~\textrm{kg/m}^{3}$ is that in this density region the tabulated EOSs deviate from the polytropic EOS for the low density region too much).
It is also shown in this figure that in the stable neutron star
sequence, when the stellar mass smaller than about $ 0.2~
M_{\odot}$, the radius will increase rapidly as the stellar mass
decreases. Before the stellar mass reaches the $M_{min}$, the
radius of neutron star can increase over 200 km, which is far
larger than the size of the normal neutron stars, this point can
be seen more clearly in Fig. \ref{5}. The larger size is the main
characteristics of the neutron star with mass around the
$M_{min}$.

\begin{figure}\label{6}
\centering
\includegraphics [width=0.6\textwidth]{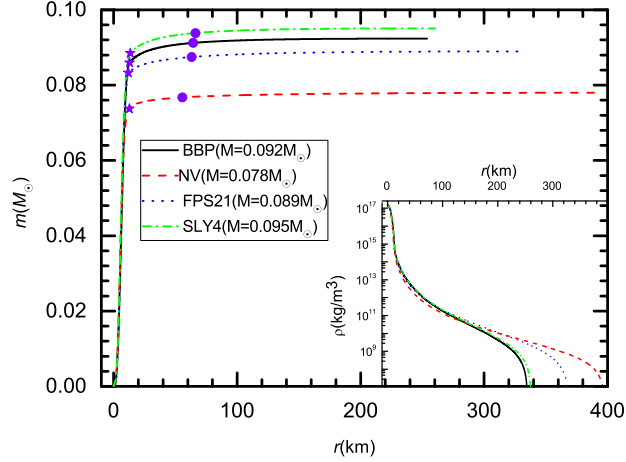}
\caption{\label{6} The mass profile for the neutron star with
minimum mass, where the stars mark the points with density at
neutron-drip point and the circles mark the points at density
$10^{12}~kg/m^3$. The inset is the corresponding density profile.
}
\end{figure}

\begin{figure}\label{6add}
\centering
\includegraphics [width=0.6\textwidth]{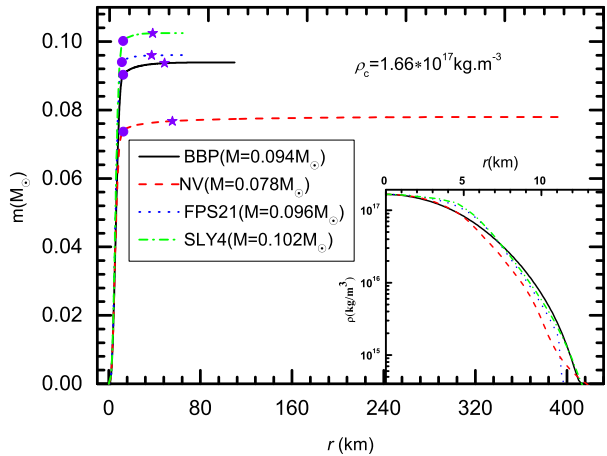}
\caption{\label{6add} The mass profile for the neutron star at
central density $1.66\times 10^{17}kg/m^3$, where the stars mark
the points with density at neutron-drip point and the circles mark
the points at density $10^{12}~kg/m^3$. The inset is the  density
profile for the inner core with density $> 4\times
10^{14}~kg/m^3$. }
\end{figure}

\begin{table}
\caption{\label{tab:table1}the properties of non-rotating neutron
stars at minimum mass points }
\begin{ruledtabular}
\begin{tabular}{cccccccr}
& EOS & $M/M_\odot$  & $R$ $(km)$ & $\rho_c$ $(10^{17}kg/m^{3})$ & $P$ $(s)$ & $I$ $(10^{37}kg*m^{2})$ & $Z$ $(10^{-4})$$$\\
\hline
& BBP & 0.092 & 254 & 1.466 & 0.230 & 2.53 & 5.37\\
\hline
& NV & 0.078 & 394 & 1.660 & 0.483 & 3.39 & 2.92\\
\hline
& FPS21 & 0.089 & 330 & 1.493 & 0.346 & 3.56 & 3.98\\
\hline
& SLY4 & 0.095 & 261 & 1.501 & 0.236 & 2.80 & 5.38\\
 \end{tabular}
\end{ruledtabular}
\end{table}

\begin{table}
\caption{\label{tab:table2} the properties of non-rotating neutron
stars at the crust-core transition density $\rho_t$ (For SLY4 EOS,
the star with $\rho_c=\rho_t$ is unstable)}
\begin{ruledtabular}
\begin{tabular}{cccccccr}
& EOS & $M/M_\odot$  & $R$ $(km)$ & $\rho_c$ $(10^{17}kg/m^{3})$ & $P$ $(ms)$ & $I$ $(10^{36}kg*m^{2})$ & $Z$ $(10^{-3})$ \\
\hline
& BBP & 0.100 &56.2 & 2.0 & 23.0 & 5.10 & 2.60\\
\hline
& NV & 0.091 & 42.4 &2.0 & 15.8 & 3.49 & 3.10\\
\hline
& FPS21 &0.106 & 40.7 & 1.8 & 13.8 &4.67 & 3.91\\
 \end{tabular}
\end{ruledtabular}
\end{table}

 Based on the TOV equations, one EOS of dense matter
determines  a unique sequence of mass-radius of neuron stars. That
is,  the solutions provide a unique map between the
 pressure-density relation $P(\rho)$ and the
mass-radius $M(R)$ relation of stars \cite{Lindblom1992}. This
unique mapping provides a useful way to infer the  EOS of dense
matter from astrophysical observations of a set of neutron stars
with both mass and radius observed-values.  The measurement of the
low-mass neutron stars in the future is a key factor for the
entire mass-radius relation if we want to constrain the EOS very
well by this way. As it has been mentioned in Sec. II, the EOS of
the inner crust matter is very difficult to be determined by the
terrestrial experiments because of the  gas of dripped neutrons,
thus the astrophysical measurements for the low-mass neutron stars
become very important   to accurately determine the EOS of the
dense matter, including that of the inner crust. It has been
turned out that EOSs have certain characteristics that make it
possible to qualitatively infer
 the stellar properties from some special density
region.  For the normal neutron stars, it is found that
 the maximum mass is determined primarily by the
behavior of  EOS at the highest densities (about $7\sim~8
\rho_{0}$ , where $\rho_{0}$ is the saturation density) and the
slope of the mass-radius relation depends mostly on the pressure
at $ \sim~4 \rho_{0}$  \cite{Ozel2009}; while the radius depends
primarily on the pressure at  $ \sim~2 \rho_{0}$
\cite{Lattimer2001}.
 Encouraged by these interesting works, as a first step, we will primarily
investigate the properties of the low-mass neutron stars, which
may be helpful for us in theoretically understanding the map
between $P(\rho)$ and $M(R)$  at the low-mass part for neutron
stars.

 The  property values of neutron stars at $M_{min}$ points are presented in
 Tab. \ref{tab:table1}. It is shown that three of the EOSs (BBS, FPS21, SLY4)
 give
 a $M_{min}$ around $ 0.09 ~M_{\odot}$, while NV EOS gives a
 relatively
 smaller minimum mass with $M_{min}= 0.078 ~M_{\odot}$.
 The difference between the
 NV EOS and other three EOSs may provide a clue to understand the
 effect of softness of  crust EOS  on the properties of the low-mass neutron stars.
 As the central densities of the neutron stars at $M_{min}$
 points
  are  below the saturation density of nuclear matter (as shown in Tab.  \ref{tab:table1}), so the properties of the
  stars at
$M_{min}$  points are only sensitive to the EOSs  at subnuclear
densities.
  In order to show the properties of the low-mass stable neutron stars around minimum mass point more clearly,
the $M-R$ relations and $M-\rho_{c}$ relations (the inset) around
the $M_{min}$ points are  plotted in Fig. \ref{5}. It is shown
that among the four adopted EOSs, the NV EOS, which is relative
softer in the density region $10^{14}~\textrm{kg.m}^{-3}$ $\sim$ $
10^{15}~\textrm{kg.m}^{-3}$ and $10^{16}~\textrm{kg.m}^{-3}$
$\sim$ $ 10^{17}~\textrm{kg.m}^{-3}$, has a higher central
density, a smaller  $M_{min}$ and a larger radius at the $M_{min}$
point.

For analyzing the properties of the neutron star around the
$M_{min}$ points, we plot the mass profiles of neutron stars at
 $M_{min}$ points  in Fig. \ref{6}, where the stars mark the points with density at
neutron-drip point and the circles mark the points at density
$10^{12}~kg/m^3$, and the inset is the density profile. It is
shown  that for the stable neutron star at minimum mass point,
there is a compact $'$core$'$ concentrated about 95\% stellar mass
in the inner core with radius within 13 km and density higher than
the neutron-drip point ($4.3\times 10^{14} ~\textrm{kg/m}^{3}$).
Another representative density point ($10^{12}~kg/m^3$) is also
marked in Fig. \ref{6}, where the mean radius $r$ is about 60 km
and the mass $m(r)$ is about 99\% of $M(R)$. These results show
that for the minimum-mass stable neutron star, most of the stellar
mass is composed of the crystal lattice of neutron-proton clusters
immersed in electron gas and is located in a small space in the
center, while  most of the stellar space is fulled of low dense
matter. These properties are totally different from the normal
neutron stars, where there is only a very thin outer crust
($\rho<4.3\times 10^{14}$). In fact, for both of the normal
neutron stars and white dwarfs, there is no such a compact
$'$core$'$ like the  minimum-mass stable neutron star, as shown in
Fig. 3.11 and Fig. 3.12  in Ref. \cite{Glendenning2000}.

 Why a softer EOS will give a lower  $M_{min}$? In what
follows we will give a qualitative discussion. For low-mass
neutron star, it is reasonable to ignore the general relativistic
effect. Thus the TOV pressure equation can be approximated as
\begin{equation}\label{II}
\frac{dp}{dr}=\frac{\rho m(r)}{r^{2}},
\end{equation}
which is just the hydrostatic equilibrium equation in Newtonian
gravity. Near the central, fixed $r$ and $dr$ (for example,
$r$=10m, $dr$=1m), a higher central density  will give a higher
$\rho$ and thus a larger $m(r)$ ($\approx 4/3\pi r^{3}\rho$),
which leads to a larger $dp$. For a softer EOS (e.g. NV EOS), a
larger $dp$ corresponds with a larger $d\rho$, which leads to a
faster density decrease. For example, at $r$=10 km,
$\rho_{BBP}$=8.2 $\times10^{15}~\textrm{kg.m}^{-3}$,
$\rho_{NV}$=3.5 $\times10^{15}~\textrm{kg.m}^{-3}$,
$\rho_{FPS21}$=6.1 $\times10^{15}~\textrm{kg.m}^{-3}$,
$\rho_{SLY4}$=6.8 $\times10^{15}~\textrm{kg.m}^{-3}$. It is clear
that the softer NV EOS has a faster density decrease.  In fact, in
the inner 10 km, the mass $m(r)$ occupies about 90\% of the the
total stellar mass. So it is easy to understand that why a softer
EOS  for the inner crust gives a smaller $M_{min}$. In order to
address this point in another way,  we plot the mass profile and
the density profile (the inset, only with density $> 4\times
10^{14}~kg/m^3$) at a fixed central density ($1.66\times
10^{17}kg/m^3$, which is the central density of NV EOS at its
minimum-mass point) in Fig. \ref{6add}. It is shown that  at the
same central density, the softer NV EOS gives a lower stellar
mass. In the inset, it is clear shown that the softer NV EOS has a
faster density decrease in the inner core.

There is another interesting phenomenon that the density profiles
(see the inset   in Fig. \ref{6}) show distinct difference at
density lower than $  10^{11}~\textrm{kg.m}^{-3}$ while the EOSs
at this density region are nearly the same (see the inset in Fig.
\ref{2}). In fact, this difference does not come from the
difference of the low density region, but comes from the mass
$m(r)$ distribution inner the star with radius of $r$. As 99$\%$
of stellar mass is distributed in the inner part where density is
higher than $  10^{11}~\textrm{kg.m}^{-3}$, so we can replace
$m(r)$ by the total stellar mass $M(R)$ in discussing the
structure of the low density part. For the star's outer part with
lower density ($< 10^{11}~\textrm{kg.m}^{-3}$), ignoring the
general relativistic effect,   Eq. \ref{II} can be approximated as
\begin{equation}\label{I}
\frac{dp}{dr}=\frac{\rho M(R)}{r^{2}}.
\end{equation}
As the EOSs in this density region are similar, it is easy to
understand that with a fixed $dp$, a smaller stellar mass $ M(R)$
will give a larger $dr$, and finally lead to a relatively larger
stellar radius. We can also qualitatively understand this in
another way. For a neutron star based on NV EOS, as it has a core
with lower mass, the gravitational potential at its outer crust is
thus weaker than that of a neutron star based on the other EOSs in
Fig.8, which leads to a relative incompact matter distribution and
therefore result in a larger stellar radius.

\begin{figure}
\centering
\includegraphics [width=0.6\textwidth]{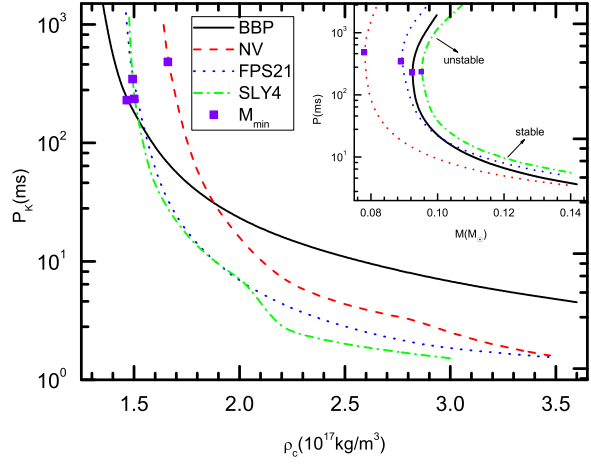}
\caption{ \label{7} The  Keplerian period ($P_{K}$) as a function
of the central density  $\rho_{c}$, where $P_{K}$ denotes the
period of a star rotating at Keplerian frequency. The inset shows
the $P_{K}$ - $M$ relations. }
\end{figure}

\begin{figure}
\centering
\includegraphics [width=0.6\textwidth]{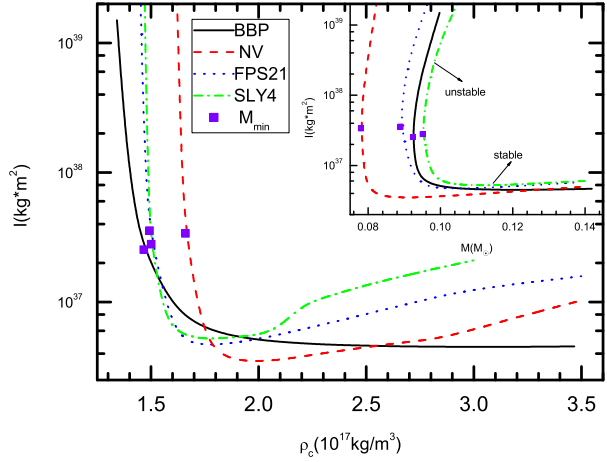}
\caption{ \label{8} The  moment of inertia $I$ as a function of
the central density $\rho_{c}$. The inset shows the  $I$ - $M$
relations. }
\end{figure}

\begin{figure}
\centering
\includegraphics [width=0.6\textwidth]{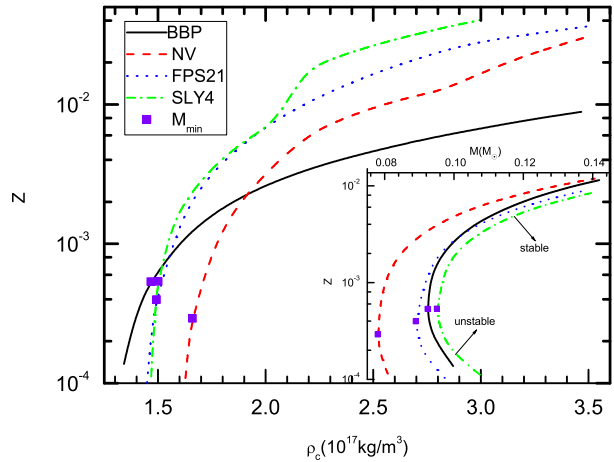}
\caption{\label{9} The  redshift $z$ as a function of the central
density $\rho_{c}$. The inset shows the  $z$ - $M$ relations. }
\end{figure}

Beside the mass-radius relation,  it is expected there are other
 observable properties for the low-mass neutron stars  can also
provide constraints on the EOSs of the inner crust matter.
 As a first step,
  we will investigate three   properties
of the low-mass neutron stars based on the currently available
  EOSs for the inner crust matter, which may be helpful in extracting the constraint on the EOSs of the crust matter in the future observations on
   the low-mass neutron
  stars. These three properties are  the spin period at Keplerian frequency, the moment of inertia and the
  redshift,  which are presented in Figs. \ref{7}-\ref{9},
  respectively. Meanwhile, in order to quantitatively show them, we also present the property parameters at  $M_{min}$ points and at crust-core
  transition densities in Tabs. \ref{tab:table1} and \ref{tab:table2},
  respectively. It is shown  that for a neutron star at minimum mass $M_{min}$ ,
  its Keplerian period is several hundred millisecond, which is beyond the period of a  millisecond
  pulsars, but still in
  the region of the observed pulsars periods. From this point of view, we can not rule out the possibility that
  some of the observed pulsars may have very lower stellar mass.
  As shown in Fig.\ref{7}, around the $M_{min}$ point, as the central density increases, the Keplerian period decreases very quickly.
    When the stellar mass increases up to about $0.1~M_\odot$, as shown in Tabs. \ref{tab:table2},
  the star can finish a spin in about 20 ms, which already lies in the period region of millisecond
  pulsars. Moreover, because of the larger size,  we can qualitatively expect  that a low-mass neutron star  around the $M_{min}$ point
  may have a notable rotation deformation under the rapid spin.

It has   been recognized that the measurement of spin-orbit
coupling provides a way to   determine the moment of inertia of a
star in a double pulsar system \cite{Lyne2004,Lattimer2005}. As
the moment of inertia of a star is close related to its mass and
radius, if the mass and the moment of inertia     are observed for
the same neutron star,  the radius can be determined very well,
and then the EOS   can be constrained strictly. As a first step,
the theoretical expected value of  the moment of inertia for the
low-mass neutrons star is studied here. Near the minimum mass
point, because of the larger size (with radius about 200 km), the
moment of inertia is in an order of $10^{37} kg\cdot m^{2}$, as
shown in Tab. \ref{tab:table1} and Fig. \ref{8}. Near the minimum
mass point, due to the rapid decrease of radius as the central
density increasing, the moment of inertia decreases very quickly.
At a stellar mass about $0.1~M_\odot$, the moment of inertia is in
an order of $10^{36} kg\cdot m^{2}$, as shown in Tab.
\ref{tab:table2}. By the way, for a normal neutron star with mass
of $1.4~M_\odot$, its moment of inertia is in an order of $10^{38}
kg\cdot m^{2}$ \cite{Fattoyev2010,Wen2011}.

The measurement of redshift provides another way to constrain the
radius and EOS of compact stars \cite{Cottam2002}. For a low-mass
neutron star near the minimum mass point, because of the smaller
mass and the larger size, the surface gravitational redshift is
very tiny, with an order of $10^{-4}$, as shown in Tab.
\ref{tab:table1}, which is approximately three orders of magnitude
less than that of the normal neutron star
\cite{Cottam2002,Wen2011}. This result reminds us that if a very
tiny reshift is observed related to a pulsar in the future, it is
a clue that there is a low-mass neutrons star.  It is also because
of the rapid decrease of radius as the central density increasing,
  the redshift has a two orders of
magnitude increase in the mass region from the minimum mass point
to about $0.2~M_\odot$, as shown in   the inset of Fig.
 \ref{9}.

In the end, we present a brief discussion for the rotation effect
  on the minimum mass and radius of neutron star. Theoretically,
   at a fixed central density, a lower mass neutron star has a
  slighter rotation effect on both the stellar mass and radius at its Kepler frequency, as the
  lower mass neutron star only can support a slower Kepler
  frequency, and thus only causes a smaller centrifugal force  \cite{Wen2011,Cook1994}. The fastest observed spin
  frequency
  of pulsar is 716 Hz (corresponding with a period of 1.4 ms) \cite{Hessels2006}. For the FPS21
  EOS, in order to support such a spin frequency, its stellar mass
  must be larger than about 0.8 $M_{\odot}$  (See Fig. 7 in Ref. \cite{Wen2011}, where the FPS21 EOS is denoted as
  APR). Therefore, it is easy to understand that when   the rotation effect is taken account of, the minimum mass of  stable neutron star supported by
  an
  EOS is   determined by the fast spin frequency (no longer the lowest
  point in the mass-radius sequence). For the detailed discussion for
  the rotation effect on the minimal mass and radius of neutron
  star, please refer to Ref. \cite{Haensel2002}.

\section{Conclusions}
Up to date, all of the observations  still can not confirm or rule
out the existence of the  low-mass neutron stars in the universe.
Moreover, we also cannot rule out the existence of the  low-mass
neutron
 stars in theory.
 As most of the theoretical
researches focused on the normal neutron stars (with mass $ \sim
1.4 M_{\odot}$) or the massive neutron stars (with mass $ \sim 2
M_{\odot}$) and few of them concentrated on the low-mass neutron
stars (with mass $ < 0.5 M_{\odot}$), and also because that there
is a close relation between the low-mass neutron star and the EOS
of the inner crust matter, thus it is interesting to  carry out an
investigation on the low-mass neutron stars. In this paper, by
employing  four typical equation of states (EOSs) for the inner
crust, we investigate the detailed properties of the low-mass
neutron stars. Based on the well-known fact that there is a big
gap between the neutron stars and white dwarfs in the mass-radius
sequence of compact stars, according to the mass-radius relation
of the four adopted EOSs,  we conclude a rough forbidden-region
for the central density and stellar radius to form a compact
stars, that is, there is no compact star in nature having central
density in the region from about $10^{12}~\textrm{kg/m}^3$ to
$10^{17}~\textrm{kg/m}^3$, and there is also no compact star
having a radius in the region from about 400 km to 2000 km.
Limited by the EOS samples, we only conclude here a rough
forbidden-region for the compact stars. If more precise EOS or
more EOS samples for the crust matter are obtained in the future,
one can give a stricter forbidden-region.  The properties of the
low-mass neutron stars are analyzed. It is shown that for the
stable neutron star at minimum mass point, the stellar size (with
radius $>$ 200 km) is much larger than that of the normal neutron
stars (with radius about 10 km), and there is a compact $'$core$'$
concentrated about 95\% stellar mass in the inner core with radius
within 13 km and density higher than the neutron-drip point
($4.3\times 10^{14} ~\textrm{kg/m}^{3}$). This property is totally
different from that of the normal neutron stars and white dwarfs.

Three
  quantities of the low-mass neutron stars (the spin period at Keplerian frequency, the moment of inertia and the
  redshift),   are calculated and discussed. It is found that
 for a stable neutron star with stellar mass near the
minimum mass, the Keplerian period is several hundred millisecond,
which is beyond the period of a millisecond pulsars, but still in
the region of the observed pulsars periods; the moment of inertia
is in an order of $10^{37} kg\cdot m^{2}$; and the surface
gravitational redshift is in an order of $10^{-4}$.

If the new observation apparatuses find the low-mass neutron star
in the future, it will provide a way to constrain the EOSs of
nuclear matter at subnuclear densities strictly. Thus it will be
an interesting topic to investigate the inverse stellar structure
problem to determine  the equation of state of the matter at
subnuclear densities.

\begin{acknowledgements}
We would like to thank  G. C. Yong and X. D. Zhang for helpful
discussion and the referee for helpful comments and suggestions.
This work is supported by NSFC ( Nos. 11275073 and 11305063). This
project  has made use of NASA's Astrophysics Data System.

\end{acknowledgements}

%\newpage

\end{document}